\documentclass{pasj00}
\begin{document}
\SetRunningHead{Lobanov et al.}{VSOP imaging of S5~0836$+$710}
\Received{2000/12/31}%{yyyy/mm/dd}
\Accepted{2001/01/01}%{yyyy/mm/dd}

\title{Dual frequency VSOP imaging of the jet in S5~0836$+$710}

%%% begin:list of authors
\author{Andrei \textsc{Lobanov}, Thomas \textsc{Krichbaum}, Arno
  \textsc{Witzel} and J. Anton \textsc{Zensus}%
  }
\affil{Max-Planck-Institut f\"ur Radioastronomie, Auf dem H\"ugel 69, 53121}
\email{alobanov@mpifr-bonn.mpg.de}

%%% end:list of authors

%%% Please use the following style in case that sorting by 
%%% affiliation is impossible. 
%
% \author{%
%   D-Firstname \textsc{D-Familyname}\altaffilmark{1}
%   E-Firstname \textsc{E-Familyname}\altaffilmark{1,2}
%   and
%   F-Firstname \textsc{F-Familyname}\altaffilmark{2}}
% \altaffiltext{1}{Address of Institute}
% \email{ddddd@xxx.xxx.xx.xx}
% \email{eeeee@xxx.xxx.xx.xx}
% \altaffiltext{2}{Address of Institute}

%% `\KeyWords{}' always has to be placed before `\maketitle'.
 %Do NOT move this preamble from here!

\maketitle

\begin{abstract}
  The luminous high-redshift ($z=2.17$) quasar S5~0836$+$710 has been observed
  in October 1997 with the VSOP at 1.6\,GHz and 5\,GHz. We report here a
  previously unpublished image made from the data at 1.6\,GHz and compare the
  structure of a relativistic jet in this quasazr at the two frequencies.  We
  present a spectral index image tracing spectral properties of the jet up to
  $\sim 40$ milliarcsecond distance from the nucleus. The curved jet ridge
  line observed in the images and the spectral index distribution can be
  described by Kelvin-Helmholtz instability developing in a relativistic
  outflow with a Mach number of $\sim 6$. In this description, the overall
  ridge line of the jet is formed by the helical surface mode of
  Kelvin-Helmholtz instability, while areas of flatter spectral index embedded
  into the flow correspond to pressure enhancements produced by the elliptical
  surface mode of the instability. An alternative explanation involving a
  sequence of slowly dissipating shocks cannot be ruled out at this point.
\end{abstract}

\section{Introduction}

The VSOP (VLBI\footnote{Very Long Baseline Interferometry} Space Observatory
Program) is the first Space VLBI (SVLBI) mission that has been running in a
facility mode. VSOP observations have utilized a worldwide array of radio
telescopes and an orbiting 8-meter antenna deployed on the Japanese satellite
HALCA\footnote{Highly Advanced Laboratory for Communication and Astronomy}
(\cite{hir96}).  It has been demonstrated that the ground-space baselines
provided by HALCA contain a unique information about source structure, and
that this information cannot be obtained by super-resolution applied to ground
VLBI images \citep{lob+99,mur99}. Regular VSOP observations started in
September 1997 at 1.6 and 5\,GHz, quickly turning it to an excellent tool to
study the morphology extragalactic flows {\em en mass} (\cite{sco+04}, and
references therein) and on a case by case basis (e.g.,
\cite{hem00,pin+99,bb00,jin+01,lob+01,lz01}).  VSOP observations at 1.6\,GHz
match closely the resolution of the VLBA\footnote{Very Long Baseline Array of
  the National Radio Astronomy Observatory, USA} array at 5\,GHz
(\cite{zdn95}, and references therein). This can be used effectively for
studying the distribution of spectral index in extragalactic jets
\citep{edw+01}.  The quasar 0836$+$710 (4C71.07, z$=$2.17) was observed with
the VSOP in 1997, on October 3 (at 1.6\,GHz) and October 7 (at 5\,GHz).  The
observation at 5\,GHz was presented in \citet{lob+98}.

0836$+$710 is an ultra-luminous quasar showing correlated (Otterbein et al.
1998) broad-band variability in gamma-ray (\cite{fic94}), X-ray
(\cite{bru94}), optical (\cite{lin+93,vil+97}), mm- and cm- radio regimes
\citep{mb94}.  The source has a well collimated VLBI-scale jet
\citep{hcd01,zen+02} extending out to $\sim$240 milliarcseconds \citep{hum+92}.
The jet shows typical proper motions of 0.2--0.3\,mas/yr \citep{ott+98,jor+01}
and it is substantially curved, with lateral displacements of its ridge line
and oscillations of its transverse width (possibly correlated with the
observed jet speeds; \cite{kri+90}).

We present here a 1.1-mas resolution VSOP image of 0836$+$710 made at
1.6\,GHz, and discuss the morphological and spectral properties of the compact
jet in this quasar.  For the purpose of facilitating comparisons with previous
measurements, the big bang cosmology with the Hubble constant $H_0 =
100\,h$\,km\,s$^{-1}$\,Mpc$^{-1}$ and deceleration parameter $q_0 = 0.5$ is
adopted throughout the paper.  For 0836$+$710, this translates into a linear
scale of 4.0\,pc/mas and corresponds to apparent speed of 41.6\,$c$ for an
observed proper motion of 1\,mas/yr.

\section{VSOP observations and data reduction}

\begin{figure}
  \begin{center}
    \FigureFile(80mm,80mm){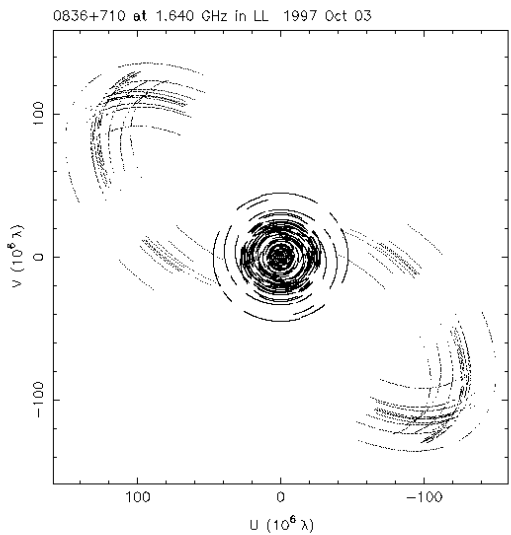}
  \end{center}
 \caption{\label{lobanov-fig1} The {\it uv}-coverage the VSOP
 observation of 0836$+$710. The inner circles correspond to the
 baselines between ground radio telescopes; the space baselines are
 grouped into larger ellipses extending at a P.A.$\approx 45^{\circ}$.}
\end{figure}

0836$+$710 was observed with the VSOP at 1.6\,GHz for 11.5 hours with the VLBA
providing ground support for the observation.  The data were recorded in the
VLBA format, using total observing bandwidth of 32\,MHz divided in two
intermediate frequency (IF) bands, each having 256 spectral channels.  The
data were correlated at the VLBA correlator in Socorro, with output
pre-averaging time of 1.966 and 0.524 seconds for the ground and space
baselines respectively.  Data from HALCA were received for the total of 5
hours at the satellite data acquisition stations in Tidbinbilla (Australia),
Robledo (Spain), and Green Bank (USA). Fringe visibilities were detected in
the HALCA data recorded at all three STS. The RMS noise on the HALCA baselines
is about 4.5 times higher than on the ground baselines.  The {\em uv}-coverage
of the final correlated dataset is shown in Fig.~\ref{lobanov-fig1}.  The
minimum {\em uv}-spacing of the data is 0.77\,M$\lambda$, corresponding to an
angular scale of 260\,mas. The maximum {\em uv}-spacings of the ground and
space baselines are 47\,M$\lambda$ and 180\,M$\lambda$, respectively. These
values indicate that the VSOP data provide an improvement of resolution by a
factor of $\sim$3.5 compared to ground VLBI observations at the same
frequency. The estimated thermal RMS noise of the VSOP data is 0.75\,mJy/beam,
corresponding to a brightness temperature of $1.1\times 10^6$\,K.

\subsection {Fringe fitting}

The correlated data were fringe-fit and imaged using AIPS\footnote{The NRAO
  Astronomical Image Processing System} and DIFMAP (\cite{she+94}).  Antenna
gain and system temperature measurements were applied to calibrate the
visibility amplitudes. After inspecting the IF bandpasses, the last 46
channels were flagged in each IF, owing to significant (50-80\%) amplitude
reduction. This has reduced the total observing bandwidth to 26.2 MHz. The
recorded phase-calibration signals from the observing log were applied to the
VLBA antennas. The residual delays and rates were corrected using the
single-band (SB) and multi-band (MB) delay fringe fitting. We used solution
intervals of 3 minutes (SB) and 5 (MB) minutes, and accepted all solutions
with ${\rm SNR}>7$.  After the fringe fitting, the residual phase variations
were found to be within 2$^{\circ}$ on the ground baselines, and $<10^{\circ}$ on the
space baselines. We then averaged over all frequency channels, and calibrated
the phases with a point source model (to enable time averaging).

\subsection{Self-calibration and imaging}

\begin{figure*}
  \begin{center}
    \FigureFile(160mm,130mm){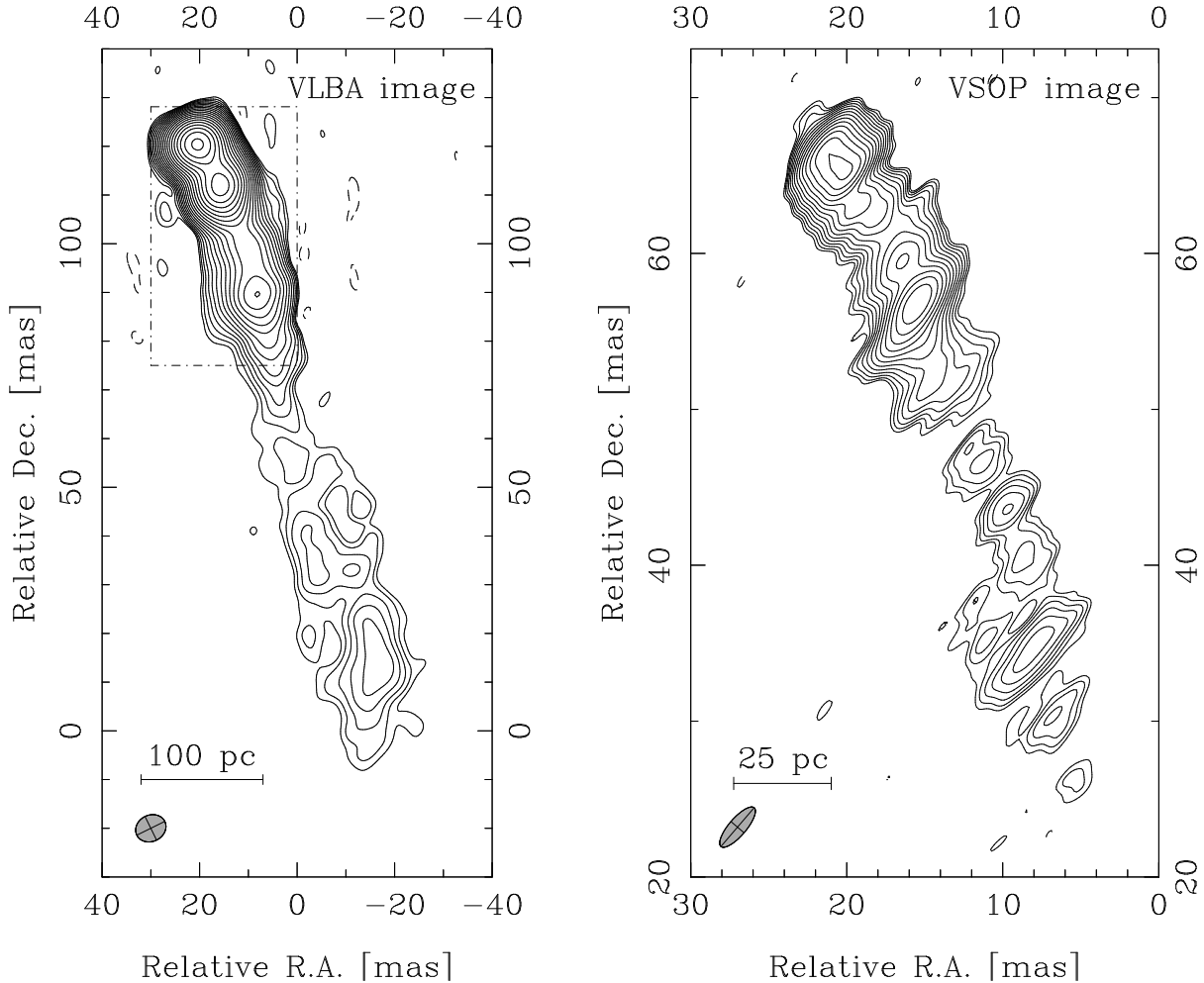}
  \end{center}
\caption{
  Ground array VLBA image (left) and VSOP image (right) of 0836$+$710 at
  1.6\,GHz. In the VLBA image, the dotted-line rectangle shows the area
  covered by the VSOP image. The image parameters are summarized in Table~1.
  Contours are drawn at $-1,\,1,\,\sqrt{2},\,2,\,...$ of the respective lowest
  contour levels, $S_\mathrm{l}$ listed in Table~1. The respective synthesised
  beams and linear scales are drawn in the bottom-left corner of each image.
  The synthesised beam in the VLBA image is $6.4\times5.4$\,mas at a P.A. of
  $-63\,.\!\!\!{}^{\circ}9$. The synthesised beam in the VSOP image is
  $3.3\times 1.1$\,mas at a P.A. of $-40\,.\!\!\!{}^{\circ}6$.
\label{lobanov-fig2}
}
\end{figure*}

After fringe fitting, the data were exported into DIFMAP and further
time-averaged into 60-second bins (with the resulting spatial dynamic range of
$\approx 190$ and the largest detectable scale of $\approx 270$\,mas). The
amplitude and phase errors were calculated from the scatter in the un-averaged
data.  The hybrid imaging using the CLEAN algorithm was applied to produce
final, self-calibrated images. Both phase and amplitude self-calibration have
been applied, with amplitudes being allowed to vary only after the total model
flux has approached the zero spacing flux to within 3\%.

\begin{table}
\caption{Parameters of the images in Fig.~1}
\label{lobanov-tb1}
\begin{center}
\begin{tabular}{l|rrrrrr}\hline\hline
Image & $S_\mathrm{t}$ & $S_\mathrm{p}$ & $S_\mathrm{l}$ &$S_\mathrm{n}$ &
      $\sigma_\mathrm{r}$ & $D_\mathrm{r}$ \\ \hline
VLBA  & 3950 & 1580 & 2.0   & 3.2 & 0.6  & 6500:1 \\
VSOP  & 3780 & 600  & 6.0   & 7.0 & 1.5  & 2500:1 \\ 
\hline
\end{tabular}
\end{center}
{\bf Notes:}~$S_\mathrm{t}$\,[mJy] -- total CLEAN flux density;
$S_\mathrm{p}$\,[mJy/beam] -- peak flux density;
$S_\mathrm{l}$\,[mJy/beam] -- lowest positive contour level in the image;
$S_\mathrm{n}$\,[mJy/beam] -- largest absolute residual flux density;
$\sigma_\mathrm{r}$\,[mJy/beam] -- residual rms noise; $D_\mathrm{r}$ --
peak-to-rms dynamic range.
\end{table}

To evaluate and compare the quality of the VSOP data to that of the ground
VLBI observations, we produce two different images: a ``VSOP image'' made from
the entire dataset and a ``VLBA image'' obtained from the visibility data
measured between the ground telescopes.  Natural weighting is applied for
gridding the ground array data, enhancing the sensitivity to extended emission
at the price of slightly decreased image resolution. The gridding weights are
also scaled by amplitude errors raised to the power of $-1$.  Uniform
weighting is applied for gridding the full-resolution VSOP data, which
provides a better angular resolution at the expense of lowering slightly the
sensitivity to extended structures.  In the VSOP data, scaling the gridding
weights by the amplitude errors weights down significantly all of the long
{\it uv}-spacings, since the noise on HALCA baselines is much higher than on
the ground baselines. To avoid the down-weighting of the HALCA baselines, the
amplitude scaling of gridding weights is switched off. As a safeguard measure,
only phase self-calibration is applied to the VSOP dataset. After a good fit
to the data is achieved, the antenna gains are adjusted, correcting for small
constant offsets between the model and the visibility amplitudes.
Fig.~\ref{lobanov-fig2} presents the VLBA image and a the VSOP image of
0836$+$710 obtained using this procedure.  Table~\ref{lobanov-tb1} lists basic
parameters of the two images.  

The quality of the images can be assessed from
the residual rms, $\sigma_\mathrm{r}$, and the maximum amplitude,
$S_\mathrm{r}$, in the residual image. Assuming that the ideal, theoretical
noise is Gaussian, with a zero mean, the measured
$\sigma_\mathrm{r}$ can be used to estimate the expected value of $S_\mathrm{r}$ in a
residual image
\[
|S_\mathrm{r,exp}| = \sigma_\mathrm{r} \left[ \sqrt{2} \ln \left(
\frac{N_\mathrm{pix}}{\sqrt{2\pi} \sigma_\mathrm{r}} \right)\right]^{1/2}\,,
\]
where $N_\mathrm{pix}$ is the total number of pixels in the image. For the
images of 0836$+$710, $N_\mathrm{pix} = 1024^2$, and the expected values of
$S_\mathrm{r}$ are 2.6\,mJy/beam and 6.3\,mJy/beam for the VLBA and the VSOP
image, respectively.
The ratio
$\epsilon_\mathrm{r} = S_\mathrm{r}/S_\mathrm{r,exp}$ expresses the quality of
the residual noise distribution. Ideally $\epsilon_\mathrm{r}$ should be close
to unity. Large values of $\epsilon_\mathrm{r}$ indicate that not all the
structure has been adequately recovered, while small values of
$\epsilon_\mathrm{r}$ imply that the number of degrees of freedom in the model
representation of the emission is too large. A relative deviation of the
measured noise distribution from the ideal, Gaussian noise is then given by
$\kappa_\sigma = \exp(|\ln\,\epsilon_\mathrm{r}|)-1$, in units of the Gaussian standard
deviation, $\sigma$. This yields $\kappa_\sigma(\mathrm{VLBA}) = 0.23\sigma$ and
$\kappa_\sigma(\mathrm{VSOP}) = 0.11\sigma$, indicating that the distributions
of residual noise are sufficiently close to the Gaussian distribution. This
implies that both images adequately represent the structure detected in the
visibility data.

\begin{figure*}
  \begin{center}
    \FigureFile(160mm,102mm){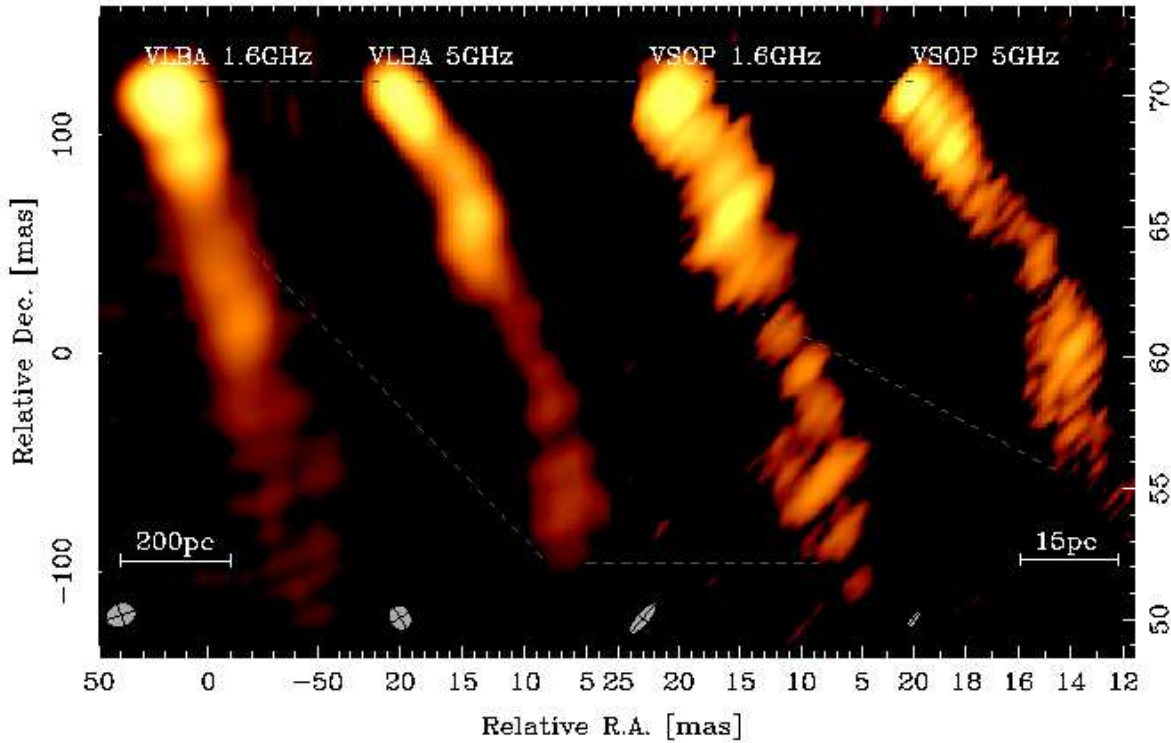}
  \end{center}
 \caption{ 
   Collage of the VLBA and VSOP images at 1.6\,GHz and 5\,GHz. Tick marks
   along the R.A. axis give relative scale and position of the emission in the
   individual images.  In each individual image, the scales of the respective
   R.A. and Dec. axis are equal.  The Declination axes are marked explicitly
   for the 1.6\,GHz VLBA image (left side) and 5\,GHz VSOP image (right side).
   Dashed lines indicate similar scales in different images. Ellipses mark the
   synthesised beams of each individual image, plotted to the respective
   scales. The 1.6\,GHz VLBA image is made with a Gaussian taper of 0.5 at
   $u\ge 10$\,M$\lambda$ to emphasize structures on scales $\ge 100$\,mas (the
   un-tapered image is shown in the left panel of Fig.~1). The synthesised beam
   of the tapered image is $14.5\times 11.2$\,mas at a P.A. of
   $-71.\!\!\!^{\circ}$. Formal parameters and a detailed discussion of the
   images at 5\,GHz are given in \citet{lob+98}. The VLBA image at 5\,GHz and
   the VSOP image at 1.6\,GHz have comparable synthesised beams due to roughly
   matching {\em uv}-coverages. These images can be used for mapping the
   spectral index distribution in the jet.
\label{lobanov-fig3}}
\end{figure*}

\section{Morphology and spectrum of the jet}

In both images in Fig.~\ref{lobanov-fig2}, the jet is continuous and curved,
with several enhanced emission regions embedded in it.  The apparent curvature
of the jet may reflect the presence of plasma instabilities in the jet (e.g.
MHD or Kelvin-Helmholtz instabilities). Note the roughly two times smaller
dynamic range of the VSOP image (2500:1, compared to 6500:1 obtained in the
VLBA image) caused by the poorer sampling and increased noise of the data on
the space baselines. Such a reduction of the dynamic range is typical for all
VSOP observations. One can compare the result above with the VSOP observation
of 0836$+$710 at 5\,GHz \citep{lob+98} which yielded the dynamic ranges of the
ground and space images at 5\,GHz are 4600:1 and 900:1, respectively. 

In Fig.~\ref{lobanov-fig3}, the images at 1.6\,GHz and 5\,GHz are compared,
illustrating the large magnitude of angular scales probed by ground and space
VLBI data. The jet oscillates with respect to the general direction of the jet
axis at a $\mathrm{P.A.} \approx 162^{\circ}$. In the heavily tapered VLBA
image at 1.6\,GHz, the jet is traced up to $\sim 240$\,mas
(960\,$h^{-1}$\,pc), similar to the extent of the emission recovered in the
previous VLBI$+$MERLIN observations of 0836$+$710 \citep{hum+92}.  The VSOP
image at 5\,GHz \citep{lob+98} shows the fine structure on scales down to
$\approx 0.8\,h^{-1}$\,pc.  The VLBA image at 5\,GHz has a resolving beam of
$2.15\times 1.75$\,mas that is similar to the synthesised beam of the VSOP
image at 1.6\,GHz. The VSOP beam is more elongated, but the two beams have
almost identical area (2.96\,mas$^2$ and 2.85\,mas$^2$, respectively).  This
facilitates making a spectral index map from these images.

\begin{figure*}
  \begin{center}
    \FigureFile(160mm,120mm){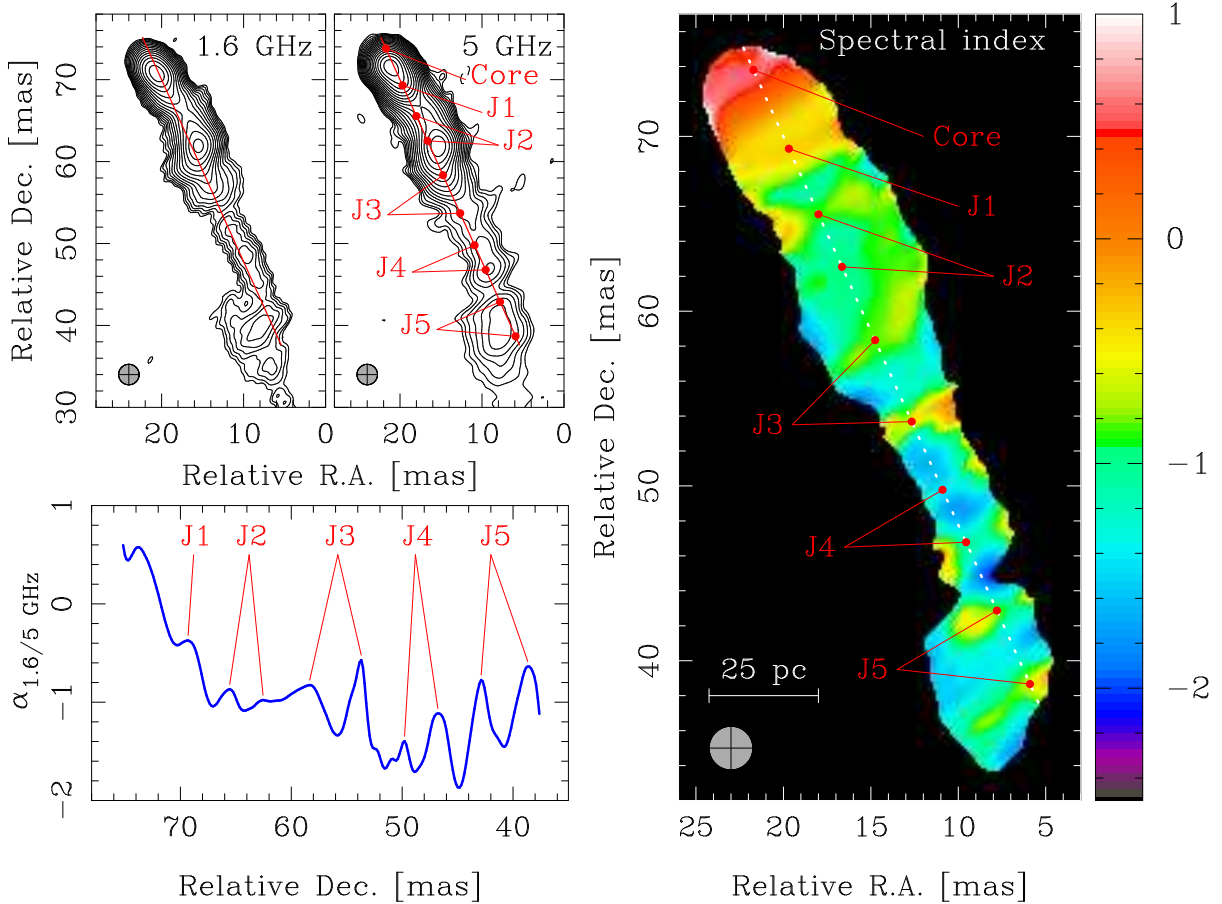}
  \end{center}
 \caption{Spectral index distribution in the jet of 0836$+$710. The
   distribution is derived from the images shown in the top left. The ground
   array image at 5\,GHz and VSOP image at 1.6\,GHz have comparable
   synthesised beams that facilitate spectral index mapping. The restoring
   beam is a circular Gaussian with a diameter of 2.5\,mas. The two images are
   aligned at the position of the peak in the region J2 (to avoid the
   frequency dependence of the apparent location of the nucleus of the jet;
   \cite{lob98a}). The resulting spectral index distribution is shown in the
   right panel. The spectral index varies smoothly along the jet and shows
   largely transverse structure indicative of shocks. The spectrum remains
   optically thick within $\approx 7$\,mas from the nucleus. Dotted line marks
   the run of a slice presented in the bottom left panel. Dots indicate the
   locations of individual peaks identified in the slice. In the optically
   thin part of the the jet, edges of all distinct regions of enhanced
   emission (marked with a letter ``J'' and successive numbers) are emphasized
   by peaks of the spectral index. The peaks are spaced at quasi-regular
   intervals, with an average separation of $4.8\pm0.2$\,mas.  This regularity
   is indicative of shocks or instabilities developing in the jet plasma.
\label{lobanov-fig4}}
\end{figure*}

\subsection{Spectral index distribution}

To derive the distribution of spectral index between 1.6\,GHz and 5\,GHz we
restore both individual images with a circular beam of 2.5\,mas in diameter
(actually increasing the beam area by a factor of $\approx 1.7$ compared to
the original synthesised beams). Natural weighting was used for producing the
image at 5\,GHz and the gridding weights were scaled by the amplitude errors
raised to the power of $-2$. The image at 1.6\,GHz was produced with the
uniform weighting and without scaling the gridding weights by the amplitude
errors.  The resulting images are shown in the upper left panels of
Fig.~\ref{lobanov-fig4} and their properties are summarized in
Table~\ref{lobanov-tb2}.

The images were aligned at the position of the peak of the brightest secondary
feature (near the outer edge of the region J2 marked in images in
Fig.~\ref{lobanov-fig4}). This measure is intended to reduce the effect of the
variable optical depth across the nuclear region. To produce the alignment,
the 5.0\,GHz image was shifted by 1.2\,mas and 1.6\,mas in right ascension and
declination, respectively. It should be noted that this shift does not
correspond directly to the core shift due to self-absorption of synchrotron
emission \citep{lob98a}, since a fraction of it comes from the shift of the
phase-center of the 1.6\,GHz VSOP image in which the peak brightness is
observed downstream from the jet base. Accurate model-fitting analysis is
required to assess the amount of core shift due to opacity in the nuclear
region. A preliminary estimate indicates a likely shift of 0.8--1.2\,mas.

After the alignment, the spectral index was calculated in all pixels with flux
densities larger than 5\,mJy/beam and 3\,mJy/beam in the 1.6\,GHz and 5\,GHz
images, respectively. The resulting spectral index image is shown in the right
panel of Fig.~\ref{lobanov-fig4}. The image shows a smooth distribution of
spectral index on angular scales of up to 40\,mas (160$h^{-1}$\,pc). The
spectral index varies from $+$0.6 in the optically thick nuclear region to
$-1.8$ in the rarefied, optically thin regions of the extended jet. Most of
the distinct features visible in the spectral index map are oriented roughly
transversely to the jet, outlining a smooth and regular evolution of the
spectral index along the flow. The optically thick emission extending up to
$\approx 7$\,mas ($\approx 30\,h^{-1}$\,pc distance from the base of the jet
is likely to mark the region where radiation from strong shocks dominates.
This scale is similar to the one found in 3C\,345 \citep{lz99}, from spectral
evolution in the regions of enhanced emission (jet components).

\begin{table}
\caption{Parameters of the single frequency images in Fig.~4}
\label{lobanov-tb2}
\begin{center}
\begin{tabular}{l|rrrrrrr}\hline\hline
Image & $S_\mathrm{t}$ & $S_\mathrm{p}$ & $S_\mathrm{l}$ &$S_\mathrm{n}$ &
      $\sigma_\mathrm{r}$ & $D_\mathrm{r}$ & $S_\mathrm{cl}$\\ \hline
1.6GHz  & 3780 & 1140 & 7.0   & 7.0 & 1.4  & 2700:1 & 5.0\\
5GHz    & 2160 & 1000 & 1.5   & 6.2 & 0.4  & 5400:1& 3.0 \\ 
\hline
\end{tabular}
\end{center}
{\bf Notes:}~$S_\mathrm{t}$\,[mJy] -- total CLEAN flux density;
$S_\mathrm{p}$\,[mJy/beam] -- peak flux density;
$S_\mathrm{l}$\,[mJy/beam] -- lowest positive contour level in the image;
$S_\mathrm{n}$\,[mJy/beam] -- largest absolute residual flux density;
$\sigma_\mathrm{r}$\,[mJy/beam] -- residual rms noise; $D_\mathrm{r}$ --
peak-to-rms dynamic range; $\S_\mathrm{cl}$ [mJy/beam] -- flux density
clipping level for the spectral index map.
\end{table}

To illustrate the connection between the spectral index distribution and the
regions of enhanced brightness embedded in the jet, a run of the spectral
index along a slice in the jet is plotted in the lower left panel of
Fig.~\ref{lobanov-fig4}.  The location of the slice is indicated in the
spectral index map and in the 5\,GHz image. The dots mark the positions of the
local peaks in the run of the spectral index. The peaks seemingly underline
the edges of the enhanced emission regions. This can be viewed as evidence for
forward and reverse shocks dissipating in the jet after injection of dense
plasma condensations at the nucleus (cf., \cite{mb94}; \cite{alo+03}). The likely
associations with the enhanced brightness regions are marked with J1--J5 in
Fig.~\ref{lobanov-fig4}.  Another possible explanation involves elliptical
modes of Kelvin-Helmholtz (K-H) instability developing in the flow
\citep{har00,lz01}. It is remarkable that the peaks in the spectral index run
are separated by similar intervals, with an average separation of $4.8\pm
0.2$\,mas. This regularity suggests that K-H instability can also play a role
in forming the observed structure. It is also possible that the observed
morphology and spectral properties result from an interplay between
relativistic shocks and instabilities \citep{lz99}. Better indications about
the nature of the extended emission should come from imaging the turnover frequency
distribution \citep{lob98b}, which is sensitive to velocity and pressure
gradients in the plasma.

\section{Discussion}

The reported observation of 0836$+$710 shows that the VSOP mission provides an
excellent opportunity for high-resolution and high dynamic range imaging of
VLBI-scale jets in extragalactic objects. The VSOP image of 0836$+$710 allows
to investigate the jet morphology and physical conditions on projected linear
scales up to 1\,kpc.  The jet is substantially curved on scales from $\sim
2$\,mas to $\sim 100$\,mas, underlining the likely complexity and possible
stratification of the flow. This is also reflected in remarkable changes of
the apparent speeds measured at different locations along the jet. Near the
core, apparent speeds $\beta_{\rm app} \approx 10\,h^{-1}\,c$ are measured. At
a 3$\,$mas core separation, the speed decreases down to $\beta_{\rm app} \sim
2$-$3\,h^{-1}\,c$, and then becomes larger once again, further out
\citep{kri+90,ott96}.

This kinematic behavior and the pronounced curvature of the jet are indicative
of Kelvin-Helmholtz (K-H) instability developing in the flow \citep{hum+92}.
Previous analysis of the jet ridge line measured in the VSOP image at 5\,GHz
(rightmost image in Fig.~\ref{lobanov-fig3}) has indicated that, on scales of
$\le 20$\,mas the jet is dominated by a helical surface (H$_\mathrm{s}$) mode
of K-H instability \citep{har00} in a relativistic flow with a Lorentz factor
$\Gamma_\mathrm{j}\approx 11$, Mach number $M_\mathrm{j}\approx 6$, and a
jet/ambient medium density ratio of $\approx 0.04$ \citep{lob+98}. With the
parameters derived in \citet{lob+98}, the H$_\mathrm{s}$ mode has a wavelength
of $\approx 7.7$\,mas, similar to the oscillations observed in the ridge line
at 5\,GHz. The corresponding wavelength of the elliptical surface mode
(E$_\mathrm{s}$) is $\approx 4.6$\,mas, which is close to the observed average
separation of $4.8\pm0.2$\,mas between the peaks in the spectral index run
(bottom left panel of Fig.~\ref{lobanov-fig4}). The instability should have a
relatively fast propagation speed $\beta_\mathrm{w} \approx 0.7\,c$. To
maintain the remarkable stability of the flow, the opening angle,
$\phi_\mathrm{j}$, of the jet should not exceed the limit of
$1/(\Gamma_\mathrm{j}\,M_\mathrm{j})$, which translates into $\phi_\mathrm{j}
\lesssim 1^{\circ}$.

The VLBA image at 1.6\,GHz (left panel of Fig.~\ref{lobanov-fig2}) gives an
hint of a structural changes at a much longer wavelength, possibly as large as
$\sim 100$\,mas. It is impossible to reproduce this wavelength in a flow
discussed above, as it would require the jet to have an opening angle far in
excess of the $1^{\circ}$ limit derived above. This implies that either this
wavelength is either produced in a flow with substantially different
properties (for instance, it could be generated in outer layers of a
stratified flow) or it is not related to K-H instability at all (it could be
due to precession or other purely kinematic factors). This is subject to a
more detailed analysis of the morphology and kinematics of the jet. At this
point, it could only be noted that such a wavelength could correspond to a
helical surface mode of K-H instability developing in outer layers of the
jet, with a transverse width about 10 times larger than the width of the flow
giving rise to the shorter wavelengths detected in the VSOP image at 5\,GHz.

Finally, the preliminary measurement of the core shift between 1.6\,GHz and
5\,GHz allow us to place limits on the parameters of the inner jet, following
the method described in \citet{lob98a}. The magnetic field strength in the
self-absorbed core should be 2\,mG and 18\,mG at 1.6\,GHz and 5\,GHz,
respectively (assuming the spectral index of $-0.7$ for synchrotron emission).
The 1.6\,GHz core should be located at about 85\,pc ($\approx 1$\,mas,
assuming a viewing angle of 3$^{\circ}$) from the central engine of the AGN.
The total luminosity of the jet reaches up to $10^{48}$\,erg/s, with about 10\%
  of that coming from the synchrotron emission. This makes 0836$+$710 one of
  the most powerful extragalactic jets certainly worth of more detailed
  studies.

\section*{Acknowledgments}

We gratefully acknowledge the VSOP Project, which is led by the Japanese
Institute of Space and Astronautical Science in cooperation with many
organizations and radio telescopes around the world.  The National Radio
Astronomy Observatory is a facility of the National Science Foundation
operated under cooperative agreement by Associated Universities, Inc.

%%%
% See the manual for the detail.
%%%

\end{document}